\documentstyle{mn}

\input epsf

\begin{document}

\def\Lya{Ly$\alpha\ $}
\def\LCDM{$\Lambda$CDM\ }
\def\HI{\hbox{H~$\rm \scriptstyle I\ $}}
\def\HII{\hbox{H~$\rm \scriptstyle II\ $}}
\def\HeII{\hbox{He~$\rm \scriptstyle II\ $}}
\def\HeIII{\hbox{He~$\rm \scriptstyle III\ $}}
\def\CIV{\hbox{C~$\rm \scriptstyle IV\ $}}
\def\SiIV{\hbox{Si~$\rm \scriptstyle IV\ $}}
\def\tauH{\tau_{\rm H}}
\def\kmsmpc{\,{\rm km\,s$^{-1}$\,Mpc$^{-1}$}\,}
\def\kel{\,{\rm K\ }}
\def\ltsima{$\; \buildrel < \over \sim \;$}
\def\lsim{\lower.5ex\hbox{\ltsima}}
\def\gtsima{$\; \buildrel > \over \sim \;$}
\def\gsim{\lower.5ex\hbox{\gtsima}}
\def\etal{{ et~al.~}}
\def\aj{AJ}
\def\apj{ApJ}
\def\apjs{ApJS}
\def\mn{MNRAS}

\journal{Preprint-04}

\title{Constraints on the ionization sources of the high redshift
intergalactic medium}

\author[A. Meiksin]{Avery Meiksin${}^{1}$ \\
${}^1$Institute for Astronomy, University of Edinburgh,
Blackford Hill, Edinburgh\ EH9\ 3HJ, UK}

%\pubyear{2004}

\maketitle

\begin{abstract}
Constraints on the ionization structure of the Intergalactic Medium
are derived as directly imposed by observations in conjunction with
the results of numerical simulations for structure formation. Under
the assumption that the population of sources dominating the UV
ionizing background at $z<6$ is the same population which reionized
the IGM, it is shown that consistency with measurements of the mean
\Lya transmitted flux at high redshifts requires the epoch of hydrogen
reionization of the IGM to have occurred at a redshift $z_{\rm
ri}<11$, {\it independent of the space density of the sources}. The
upper limit on the reionization redshift depends only on the {\it
shape} of the UV spectra of the sources. Consistency with constraints
on the reionization epoch from the {\it Wilkinson Microwave Anisotropy
Probe} requires the sources of photoionization to have had hard
spectra, such as QSOs or Population III stars. The only way to escape
these conclusions is either:\ 1.\ the sources which dominate the
photoionization background at $z<6$ are the remnants of a population
that was a much more prodigious source of ionizing photons at earlier
times, or 2.\ the sources responsible for the photoionization of the
IGM are an unknown population that contributes negligibly to the
ionization of the IGM at $z<6$. The evolution of the QSO luminosity
function over the range $3<z<6$ is estimated from recent QSO counts,
and the fraction of ionizing photons arising in QSOs is evaluated
under the assumptions of either pure luminosity evolution or pure
density evolution. It is shown that QSOs dominate the UV ionizing
background for $z<3.5$, but that it is unlikely that more than one half
to one third of the ionizing background radiation originates in QSOs
in the redshift range $4.5<z<6$. QSOs acting alone could not have
reionized the hydrogen in the IGM prior to $z\approx4$. If the QSOs
had hard spectra, however, they may have reionized the helium in the
IGM as early as $z\approx5$. The possibility that the IGM was
reionized by low luminosity AGN is discussed.
\end{abstract}

%%%%%%%%%%%%%%%%%%%%%%%%%%%%%%%%%%%%%%%%%%%%%%%%%%%%%%%%%%%%%%%%%%%%%%%%%%%%%%%

\section{Introduction} \label{sec:introduction}

The ionization structure of the Intergalactic Medium (IGM) at high
redshift has been the focus of much recent attention.  On the basis of
measurements of the mean transmitted \Lya flux in spectra of the
highest known redshift Quasi-Stellar Objects (QSOs) by the Sloan
Digital Sky Survey (SDSS), it has been suggested that the reionization
of the IGM had just completed at $z\gsim6$ (Becker \etal 2001). By
contrast, measurements of fluctuations in the Cosmic Microwave
Background (CMB) by the {\it Wilkinson Microwave Anisotropy Probe}
({\it WMAP}) suggest the IGM was reionized earlier, at $z>11$
($2\sigma$ lower limit) (Kogut \etal 2003). The most obvious candidate
sources of reionization are stars and QSOs. An assessment based on the
results of the Hubble Ultra-Deep Field (UDF) and the Great
Observatories Origins Deep Survey (GOODS) by Bunker \etal (2004)
suggests that the galaxies identified at $z\approx6$ are too few to
account for the reionization of the IGM by $z=6$. On the other hand,
Yan \& Windhorst (2004) and Stiavelli, Fall \& Panagia (2004) dissent
from this view, arguing that, with plausible assumptions, these
sources are adequate. Yan \& Windhorst further argue that galaxies
appear to be the only sources of reionization. They claim QSOs
contribute negligibly to the overall ionizing photon budget, falling
short of the required numbers by more than two orders of
magnitude. Meiksin \& White (2004) argue instead for a much smaller
discrepancy for the quasars.  The range of views in the literature
highlights the sensitivity of conclusions regarding reionization to
prevailing uncertainties, in particular the luminosity function of the
sources, their nature (especially their emergent UV spectra), and the
structure of the IGM.

A question distinct from the sources responsible for reionization is
the nature of the sources which dominate the UV photoionizing
background at $z<6$, when the IGM is clearly ionized. The populations
need not be the same, as the ionizing contribution from the sources
responsible for reionizing the IGM may decline with time, while other
sources begin to dominate the UV background of the post-reionization
universe. For instance, the mean escape fraction of ionizing photons
from galaxies at high redshifts is currently unclear, and could in
principle be sufficiently high to enable galaxies to reionize the
IGM. Steidel \etal (2001) report a detection of UV ionizing photons
from the the brightest Lyman break galaxies in a sample at
$z\approx3$, suggesting an escape fraction of about 10\%, although
this difficult measurement has not yet received independent
confirmation. Attempts to quantify the mean escape fraction have
generally resulted only in upper limits. For instance,
Fern\/andez-Soto, Lanzetta \& Chen (2002) constrain the mean escape
fraction for galaxies in the range $1.9<z<3.5$ to less that 4\%, which
may be too small for galaxies to maintain the ionization of the
IGM. Applying an escape fraction of $\sim10$\% to the full range of
galaxy luminosities, Steidel \etal obtain a value for the global
emissivity of ionizing radiation that is inconsistent with the
emissivity required by numerical simulations of the IGM to reproduce
the measured values of the mean transmitted \Lya flux at
$z\approx3$. The high emissivity obtained by Steidel \etal would
result in a much higher \Lya transmitted flux through the IGM than
measured, unless the sources were very short-lived (eg, bursting at
just $z\approx3$) (Meiksin \& White 2003). This suggests that the
escape fraction found by Steidel \etal may be anomalously high. While
an escape fraction of $1-3$\% may be just sufficient for galaxies to
maintain the ionization of the IGM at $z\approx3$, it is unclear
whether this fraction is sufficient for galaxies to meet the
requirements for reionizing the IGM (see Stiavalli \etal).

Constraints on the epoch of reionization and on the level of
ionization in the post-reionization IGM have usually been treated as
separate issues. In this paper, it is shown that the two are closely
related. Any model that meets the requirements for the reionization
epoch implies a level of ionization of the IGM just after reionization
is complete. This level must be consistent with constraints imposed by
measurements of the mean \Lya transmitted flux through the IGM. In
particular, if the high \Lya optical depth measurements at $z\approx6$
designate the tail end of the reionization epoch, then the rate of UV
ionizing photon production by the sources must \lq gracefully exit'
the reionization epoch at the level required to match the measured
\Lya optical depth measurements at $z\lsim6$.

A second purpose of this paper is to provide a statistical assessment
of the contribution of QSOs to the photoionizing background at $z>3$,
and to the overall budget of ionizing photons at $z\approx6$, using
the current best estimates of the faint (Hunt \etal 2004) and
bright (Fan \etal 2001, 2004) ends of the QSO luminosity-redshift
distribution.  Estimates of the UV background based on QSO counts are
not very well constrained at high redshifts ($z>3$) due to the low
sampling of the luminosity-redshift plane by existing surveys. The
uncertainty is sufficiently large to allow for a substantial
contribution to the UV metagalactic background from QSOs at high
redshifts (Meiksin \& White 2004).

\section{The ionization porosity of the IGM}

The epoch of reionization by sources of number density $n_S$ may be
characterised by the time at which the porosity $Q=n_S \langle
V_I\rangle$ exceeds unity, where $\langle V_I\rangle$ is the average
volume of the universe ionized per source. At earlier stages, the
filling factor of the remaining neutral gas is $f=1-Q$. For any given
evolutionary scenario of the sources, the porosity is determined by
the total rate of production ${\dot N}_S$ of ionising photons per unit
volume, which in turn scales like the total emissivity of the
sources. On the other hand, for a uniform medium, the neutral
fraction, and hence the \Lya optical depth $\tau_\alpha$, of the IGM
scales inversely with the total source emissivity. As a consequence,
the product of the two will reflect only the evolution and spectral
shape of the sources (and the properties of the IGM), but be
independent of their actual numbers. For any given source history and
spectral shape, the \Lya optical depth just following the epoch of
reionization is thus well-specified and must be consistent with
measured values.  Turning this argument around, if the evolution of
the emissivity and attenuation length are known, or reasonable guesses
made based on observations or numerical simulations, then measurements
of the \Lya optical depth may be used to provide a direct estimate of
the porosity of the IGM, and hence of the epoch of reionization.

In reality, the IGM is not uniform, but clumped into the \Lya forest,
so that the relation between the porosity and the measured mean \Lya
optical depth is indirect. (Here, the mean \Lya optical depth is
defined as $\bar\tau_\alpha=-\log\langle f_\alpha\rangle$, where
$\langle f_\alpha\rangle$ is the mean transmitted \Lya flux from a
background source.) It is necessary to appeal to models of structure
formation to establish the relation. Within the context of the models,
the intimate relation between the porosity and mean transmitted \Lya
flux persists. Setting the source emissivity at a level to reionize
the IGM at a given epoch fixes the post-reionization photoionization
rate of the IGM, and so determines the mean transmitted \Lya flux just
after reionization is completed (for any given cosmological
model). Any reionization model must yield predictions consistent with
these measured values. For instance, if the predicted photoionization
rate were too high, the predicted mean transmitted \Lya flux levels
would exceed those measured. Constraints on the reionization epoch and
the post-reionization \Lya transmitted flux values must be satisfied
simultaneously.

The results of a $\Lambda$CDM simulation run to investigate effects of
radiative transfer on the \Lya forest (Meiksin \& White 2004) are used
to quantify the emissivity required to maintain the ionization of the
IGM over the redshift range $3<z<6$ at a level consistent with current
measurements of the mean transmitted \Lya flux. The simulation was run
using a pure Particle Mesh (PM) dark matter code, and it was assumed
the gas and dark matter have the same spatial distribution, which is
found to give an adequate description of the structure of the IGM
(Meiksin \& White 2001, 2003). The parameters used for the simulation
are $\Omega_{\rm M}=0.30$, $\Omega_\Lambda=0.70$, $\Omega_b=0.045$,
$h=H_0/100$\kmsmpc$=0.70$, $\sigma_8=0.92$ and slope of the primordial
density perturbation power spectrum $n=0.95$. These values will be
used throughout. The model is consistent with existing large-scale
structure, \Lya forest flux distribution, cluster abundance and {\it
WMAP} constraints (Meiksin \& White 2004). The simulation was run
using $512^3$ particles and a $1024^3$ force mesh, in a cubic box with
(comoving) side length $25\,h^{-1}$Mpc, adequate for obtaining
converged estimates of the \Lya pixel flux distribution, the photon
attenuation length and the required mean emissivity at the Lyman edge
(Meiksin \& White 2003, 2004). A description of the parallel PM code
is given in Meiksin \& White (2003).

\begin{figure}
\begin{center}
\leavevmode 
\epsfxsize=3.3in
\epsfysize=3.3in
\epsfbox{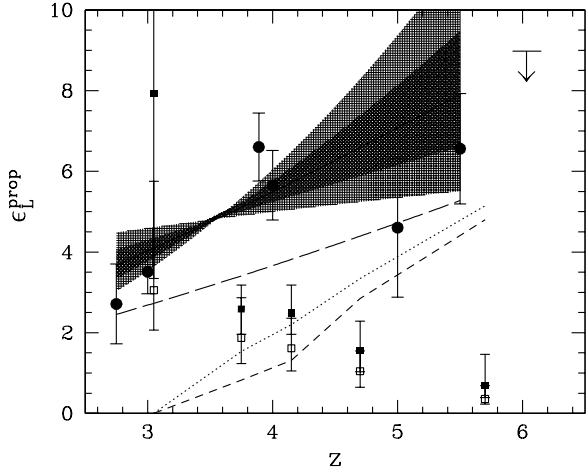}
\end{center}
\caption{Metagalactic emissivity (proper) at the Lyman edge (in units
of ${\rm 10^{26}\,erg\,s^{-1}\,Hz^{-1}\,Mpc^{-3}}$), as a function of
redshift. The data points (filled circles and upper limit point) are
the emissivity values required in a $\Lambda$CDM model of the IGM to
match the measured values of the transmitted \Lya flux through the
IGM. The shaded regions show the $1\sigma$ (dark) and $2\sigma$
(light) ranges for a power-law fit (solid line) to the evolution in
emissivity. The long dashed line shows the required source emissivity,
allowing for a boost in the total emissivity by 50\% from diffuse
emission by the IGM. The squares show the estimated contribution from
QSOs assuming either a pure luminosity evolution model (filled
squares) or a pure density evolution model (open squares) for the QSO
luminosity function. (See \S~\ref{sec:QSOs}.) The short dashed
and dotted lines show the deficit left by these (respective) QSO
models from the required source emissivity (long dashed line). The
deficit must be made up by sources other than the QSOs. (A metagalactic
spectral index of $\alpha_{\rm MG}=1$ has been assumed.)
}
\label{fig:emiss}
\end{figure}

In the limit of a medium with a photon attenuation length at the Lyman
edge short compared with the horizon scale, the photoionization rate
$\Gamma$ is related to the emissivity $\epsilon_L$ at the Lyman edge
by
\begin{equation}
\Gamma\approx\frac{\epsilon_L\sigma_L}{h_P(3+\alpha_{\rm MG})}r_0
\label{eq:Geps}
\end{equation}
where $\sigma_L$ is the photoeletric cross-section at the threshold
frequency $\nu_L$, $h_P$ is the Planck constant, $r_0$ is the photon
attenuation length, and the spectrum of the ionizing background is
taken to vary at frequencies above the threshold as a power-law
$\epsilon_\nu=\epsilon_L(\nu/\nu_L)^{-\alpha_{\rm MG}}$. In
Fig.~\ref{fig:emiss}, the resulting constraints on the metagalactic
emissivity using the data in Table 3 of Meiksin \& White (2004) are
shown. These data were based on $\Lambda$CDM model requirements to
match the mean transmitted \Lya flux measurements from a variety of
sources (see Meiksin \& White 2004, Appendix B). It is
emphasized that the emissivity and attenuation length are solved for
self-consistently in the model. Simulation-based estimates of the
required emissivity in the literature usually rely on an assumed
attenuation length (or, equivalenty, effective optical depth at the
Lyman edge) based on statistical representations of the measured \Lya forest
absorber properties (like \HI column density and Doppler parameter).
Uncertainties in the measured distributions, however,
produce uncertainties in the derived attenuation length as large as
a factor of 2--3 (Meiksin \& Madau 1993, Meiksin \& White 2003,
2004). This uncertainty translates directly into a comparable
uncertainty in the required source emissivity. For this reason, it is
crucial that the inferred emissivity from a simulation be consistent
with the attenuation length that the same simulation predicts.

A least-squares fit to the emissivity values from Meiksin \& White
(2004) over $2.75\leq z\leq5.5$ gives the comoving emissivity
\begin{equation}
\epsilon_L\approx A_{\rm MG}\left(\frac{3+\alpha_{\rm MG}}{3}\right)
(1+z)^\gamma\,h\,{\rm ergs\,s^{-1}\,Hz^{-1}\,Mpc^{-3}},
\label{eq:emiss}
\end{equation}
where $A_{\rm MG}=8.4^{+(9.7, 31)}_{-(4.5, 6.5)}\times10^{25}$ and
$\gamma=-1.6-0.6\log(A_{\rm MG}/8.4\times10^{25})$. The errors on
$A_{\rm MG}$ are the $1\sigma$ and $2\sigma$ limits. (The fit is acceptable
at the 13\% confidence level according to the $\chi^2$-test.) The upper
limit at $z\approx6$ is consistent with this fit, as shown in
Fig.~\ref{fig:emiss}.

\begin{figure}
\begin{center}
\leavevmode 
\epsfxsize=3.3in
\epsfysize=3.3in
\epsfbox{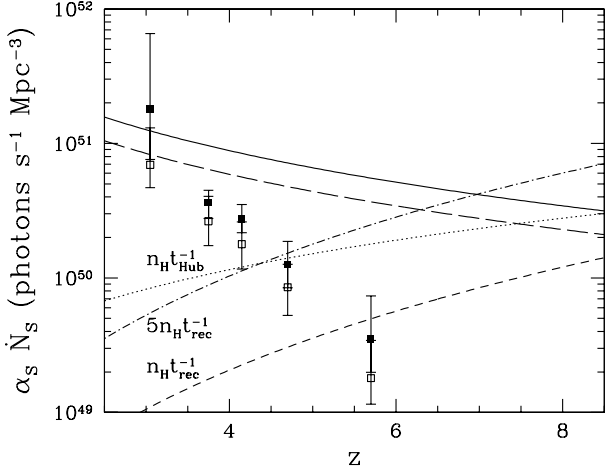}
\end{center}
\caption{The evolution of the comoving production rate of ionizing
photons.  The required rate inferred from measurements of the mean
transmitted \Lya flux is shown by the solid line. The long-dashed line
shows the required source rate if diffuse recombination radiation from
the IGM boosts the total metagalactic photoionization rate by
50\%. Also shown are the minimal ionizing photon production rate
$n_{\rm H}t_{\rm H}^{-1}$ required to ionize the IGM over a Hubble
time (dotted line), and the rate of recombinations for clumping
factors of ${\cal C}=1$ (dot-dashed line) and ${\cal C}=5$ (short
dashed line). These latter three rates are shown for $\alpha_S=1$. For
alternative values, they should be multiplied by the factor
$\alpha_S$. Also shown are the predicted contributions from QSO
sources, assuming either a pure luminosity evolution model (filled
squares) or pure density evolution model (open squares) for the QSO
luminosity function. (See \S~\ref{sec:QSOs}.)}
\label{fig:Nphot}
\end{figure}

The total rate at which ionizing photons are generated by sources, per
unit comoving volume, is related to the total (comoving) emissivity of
the sources, $\epsilon^S_L$, by
\begin{eqnarray}
{\dot N_S}&=&\int_{\nu_L}^\infty\,\frac{\epsilon^S_L}{h_P}
(\frac{\nu}{\nu_L})^{-\alpha_S}\frac{d\nu}{\nu}\approx
\frac{\epsilon_L}{h_P\alpha_S}\nonumber\\
&=&
A_S\left(\frac{3+\alpha_{\rm MG}}{3\alpha_S}\right)
(1+z)^\gamma\,h\,{\rm ph\,s^{-1}\,Mpc^{-3}},
\label{eq:Nphot}
\end{eqnarray}
where $A_S=1.3^{+(1.5, 4.6)}_{-(0.7, 1.0)}\times10^{52}$,
$\gamma=-1.6-0.6\log(A_S/1.3\times10^{52})$, $\alpha_S$ is the
spectral index of the sources at frequencies above the photoelectric
threshold ($L_\nu\propto\nu^{-\alpha_S}$), and the $1\sigma$ and
$2\sigma$ error ranges are provided. Here the approximation
$\epsilon^S_L=\epsilon_L$ has been made at the Lyman edge. This
neglects the additional diffuse emission arising from the IGM itself
which may enhance the total metagalactic photoionization rate over the
direct rate from the sources by as much as an additional 50\% (Meiksin
\& Madau 1993; Haardt \& Madau 1996). The source emissivity may thus
be overestimated by this amount. In the spirit of providing an upper
limit to the reionization redshift consistent with the measurements of
the mean transmitted \Lya flux, the more generous estimate of the
source emissivity is adopted here.

The evolution of $\alpha_S\dot N_S$ (comoving) is shown in
Fig.~\ref{fig:Nphot} (solid curve), using the fit to $\epsilon_L$
above. Allowing for a boost of 50\% in the total emissivity by diffuse
emission from the IGM reduces the required photon emission rate from
sources by the same amount (long dashed line). These rates are
compared with the minimum rate $n_{\rm H}t_{\rm H}^{-1}$ at which
ionized photons must be produced per unit volume over a Hubble time,
$t_{\rm H}\approx (2/3)(H_0\Omega_{\rm M}^{1/2})^{-1}(1+z)^{-3/2}$
(for $z>>1$), to reionize the IGM. The two rates cross at
$z\approx6-13$, depending on the value of the spectral index. This
coincidence in values is expected immediately following reionization,
and provides independent support (but not conclusive evidence) that
reionization happened during this time. The smooth evolution of the
required emissivity for $z<6$ (Fig.~\ref{fig:emiss}), reflected by the
smoothness of the fit $\dot N_S$ curve in Fig.~\ref{fig:Nphot},
suggests extrapolation to higher redshifts is reasonable. Doing so
results in too low an emission rate of ionizing photons compared with
the required minimum number at redshifts above $z\approx11$ for
$\alpha_S=0.5$, and above $z\approx8.5$ for $\alpha_S>1$.

For $z<15$, the full recombination time at the average density of the
IGM exceeds the Hubble time, and the effect of recombinations will be
small.  Because the IGM is non-uniform, however, the reionization rate
will be much higher in clumpy regions, while smaller in underdense
regions. The volume-averaged recombination time will then be reduced
by the \lq\lq clumping factor'' ${\cal C} = \langle n_{\rm
H}^2\rangle/\langle n_{\rm H} \rangle^2$ (Shapiro \& Giroux 1987;
Meiksin \& Madau 1993). Clumping will slow down the growth of
ionization regions due to recombinations within them, requiring the
recombined gas to be reionized again and so drawing on the overall
ionizing photon budget. The value of the clumping factor is
unknown. Numerical simulations prior to reionization suggest values of
${\cal C}\sim30$ at $z=8-10$ and ${\cal C}>100$ by $z=6$ (Springel \&
Hernquist 2003). It is unclear, however, that these are the
appropriate values to use. Once the gas is ionized, simulations
suggest a clumping factor of ${\cal C}\approx3-4$ (at $z=3$)
(Sokasian, Abel \& Hernquist 2003). Of particular relevance is the
number of photons per atom required to photoionize a clump (or
minihalo) of gas, which requires a detailed numerical computation
beyond the scope of current full cosmological simulations. The
computations of Shapiro, Iliev \& Raga (2004) and Iliev, Shapiro \&
Raga (2004) have begun to examine this problem. They find an average
of $2-5$ ionizing photons are required per hydrogen atom, depending on
redshift, with fewer photons required the harder the spectrum and the
fainter the mean flux incident on a minihalo.

In Fig.~\ref{fig:Nphot}, the required minimum photon production rate,
${\cal C} n_{\rm H}t_{\rm rec}^{-1}$, assuming the IGM is
recombination dominated are shown with clumping factors of ${\cal
C}=1$ and 5. Here, $t_{\rm rec}$ is the full hydrogen recombination
time at the average density of the IGM, $t_{\rm
rec}=1/n_e\alpha_B(T)$, where $n_e$ is the average electron number
density in the IGM and $\alpha_B(T)$ is the radiative recombination
rate to the excited levels of \HI. (This will be taken at the
reference temperature of $T=2\times10^4\kel$ for numerical values
below.) For ${\cal C}=5$, the recombination time is shorter than the
Hubble time for $z>4.4$, and nearly half the Hubble time at $z=7$. For
this case, recombinations will play an important role in the
reionization of the IGM. On the other hand, for ${\cal C}=3$ the
recombination time is shorter than the Hubble time only for
$z>6.6$. The role recombinations played in the reionization of
hydrogen in the IGM is therefore unclear.

Madau, Haardt \& Rees (1999) considered an IGM that was recombination
dominated, with clumping factors of up to ${\cal C}=30$. Yan \&
Windhorst based their assessment for whether a class of objects could
reionize the IGM on this value. Such a large value, however, is
inconsistent with the estimated values of $\dot N_S$ from measurements
of the mean transmitted \Lya flux in Fig.~\ref{fig:Nphot}. Scaling
the curve for $n_{\rm H}t_{\rm rec}^{-1}$ (${\cal C}=1$) up by a
factor of 30 boosts the volume-averaged recombination rate to a value
that exceeds the estimated upper limit on the source rate by a factor
of 3 at $z=6$, even allowing for a hard source spectrum of
$\alpha_S=0.5$. (The discrepancy is even larger for softer sources.)
This suggests ${\cal C}<10$ is a more realistic upper limit. (The
limit is even more restrictive for softer sources.)

The epoch of reionization is quantified using the ionization porosity
parameter. The \HII ionization porosity, $Q_{\rm HII}$, evolves
according to (Meiksin \& Madau 1993; Madau \etal 1999):
\begin{equation}
\frac{dQ_{\rm HII}}{dz}=\left[\frac{{\dot N}_S(z)}{n_{\rm H}(0)}-
\frac{{\cal C}(z) Q_{\rm HII}}{t_{\rm rec}(z)}\right]\frac{dt}{dz},
\label{eq:QHII}
\end{equation}
where $n_{\rm H}(0)$ is the comoving number density of hydrogen atoms.
If the recombination time is long compared with the Hubble time, the
recombination term in Eq.~(\ref{eq:QHII}) may be neglected.  Without a
detailed knowledge of the clumping factor and its evolution, including
feedback effects, a computation still beyond the capability of current
large-scale gravity$+$hydrodynamics$+$radiative transfer simulations,
the role of recombinations cannot be determined in detail. Here,
${\cal C}=1$ will generally be assumed, so that only an upper redshift
limit to the epoch of reionization is estimated. Significant clumping
will serve to further delay complete reionization ($Q>1$). The effect
of clumping on slowing down the reionization process may be
particularly important for \HeII reionization (Madau \& Meiksin 1994).

Using the expression Eq.~(\ref{eq:Nphot}) in Eq.~(\ref{eq:QHII}) and
integrating gives, in the absence of recombinations,
\begin{equation}
Q_{\rm HII}(z)=\frac{{\dot N}_S(z)t_{\rm H}(z)}{n_{\rm H}(0)
(1-\frac{2}{3}\gamma)}
\left[1-\left(\frac{1+z}{1+z_i}\right)^{\frac{3}{2}-\gamma}\right],
\label{eq:QHIIev}
\end{equation}
where $z_i$ is the turn-on redshift of the sources. In the presence
of recombinations (and for a redshift independent clumping factor
and IGM temperature), the solution is
\begin{eqnarray}
Q_{\rm HII}(z)&=&\frac{{\dot N}_S(z)t_{\rm H}(z)}{n_{\rm H}(0)}
\tauH^{1-\frac{2}{3}\gamma}(z)e^{\tauH(z)}\nonumber\\
&&\times\int_{\tauH(z)}^{\tauH(z_i)}d\tau\,e^{-\tau}
\frac{1}{\tau^{2-\frac{2}{3}\gamma}},
\label{eq:QHIIevrec}
\end{eqnarray}
where $\tauH(z)$ is the Hubble time measured in units of the
recombination time (including the clumping factor), $\tauH(z)={\cal
C}t_{\rm H}(z)/t_{\rm rec}(z)\simeq0.016{\cal C}(1+z)^{3/2}$, and is
equivalent to the mean number of recombinations in the IGM over a
Hubble time.  Provided $2-2\gamma/3$ is not a positive integer, the
integral may be expressed in terms of the incomplete gamma function
$\Gamma(a,x)=\int_x^\infty\,dt e^{-t}t^{a-1}$. For $\gamma>-3$
($\gamma\ne0,\, \pm\frac{3}{2}$), and $z_i\rightarrow\infty$,
\begin{eqnarray}
Q_{\rm HII}(z)&=&\frac{{\dot N}_S(z)t_{\rm H}(z)}{n_{\rm H}(0)
(1-\frac{2}{3}\gamma)}\biggl\{1+\frac{3\tauH}{2\gamma}
+\frac{3\tauH}{2\gamma(1+\frac{2}{3}\gamma)}\nonumber\\
\phantom{biggl\{}&&\times\left[\tauH
-e^{\tauH}\tauH^{-\frac{2}{3}\gamma}\Gamma\left(
2+\frac{2}{3}\gamma,\tauH\right)\right]\biggr\}.
\label{eq:QHIIevrecG}
\end{eqnarray}
When $2-2\gamma/3$ is a postive integer, the solution is
\begin{equation}
Q_{\rm HII}(z)=\frac{{\dot N}_S(z)t_{\rm H}(z)}{n_{\rm H}(0)}
e^{\tauH(z)}E_{2-2\gamma/3}(\tauH),
\label{eq:QHIIevrecE}
\end{equation}
where $E_n(x)=\int_1^\infty\,dt e^{-xt}t^{-n}$ is an exponential
integral. For the special case of a constant comoving source
emissivity ($\gamma=0$), Eq.~\ref{eq:QHIIevrecE} recovers eq.~(22) of
Madau \etal (1999). In the absence of recombinations,
$\tauH\rightarrow0$ and Eq.~\ref{eq:QHIIev} (for $z_i>>z$) is
recovered. The asymptotic series representation of
Eq.~\ref{eq:QHIIevrec} for $\tauH(z)>1$ (and $z_i>>z$) is
\begin{equation}
Q_{\rm HII}(z)\sim{\cal C}^{-1}\frac{{\dot N}_S(z)t_{\rm rec}(z)}{n_{\rm H}(0)}
\sum_{n=0}^{\infty} (-1)^n\frac{\left(2-\frac{2}{3}\gamma\right)_n}
{\tauH^n(z)},
\label{eq:QHIIevrecap}
\end{equation}
where $(a)_n\equiv\Gamma(a+n)/ \Gamma(a)=a(a+1)...(a+n-1)$. The first
term ($n=0$) is the approximation eq.~(23) of Madau \etal (1999),
valid in the limit $\tauH>>1$. Similar expressions follow for the
\HeIII porosity, $Q_{\rm HeIII}$, if $n_{\rm H}(0)$ is replaced by
$n_{\rm He}(0)$, and ${\dot N_S}$ and $t_{\rm rec}$ refer to the
production rate of \HeII ionizing photons and the recombination rate
to \HeII.

The key assumption is now made that the sources of ionization at $z<6$
are drawn from the same population as that responsible for the
reionization of the IGM. Under this assumption, the porosity $Q_{\rm
HII}$ may be estimated from Eq.~(\ref{eq:Nphot}) for ${\dot N_S}$,
extrapolated to $z>6$. This seems a reasonable assumption to make as
the inferred emissivity in Fig.~\ref{fig:emiss} evolves smoothly
over $3<z<6$ in spite of the strong downward turn in the measured mean
\Lya transmitted flux at $z>5$. To make the estimate, however, it is
first necessary to be more specific concerning the factor
$(3+\alpha_{\rm MG})/\alpha_S$ in Eq.~(\ref{eq:Nphot}).

When QSOs are the sources of reionization, the UV metagalactic
background hardens at high redshifts beyond the input spectral shape
because high energy photons travel further than photons near the
photoelectric threshold before being absorbed by the IGM. Here
$\alpha_{\rm MG}\approx1$ is adopted based on the results of Haardt \&
Madau, including re-emission from the IGM. Although the
spectral index should be computed self-consistently from the models
considered here, the uncertainty in this value introduces only a small
uncertainty in the total photoionization rate. It is noted that
$\alpha_{\rm MG}$ may be double this if young star forming galaxies
with a Salpeter IMF and solar abundances are the sources of
ionization, as they have much softer spectra than QSOs.

\begin{figure}
\begin{center}
\leavevmode 
\epsfxsize=3.3in
\epsfysize=3.3in
\epsfbox{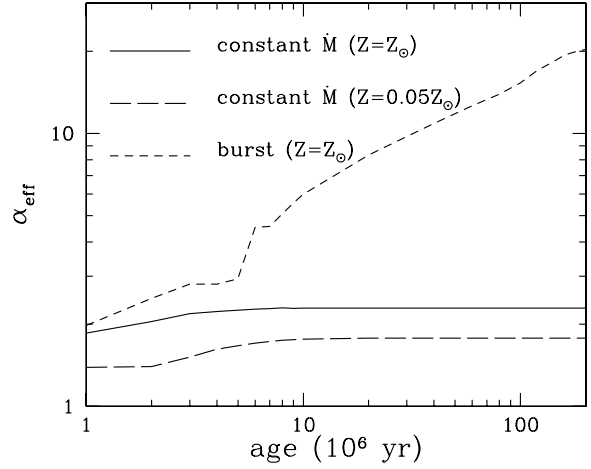}
\end{center}
\caption{The evolution of the effective spectral index $\alpha_{\rm eff}$
(see text) for a starburst of solar metallicity with a constant star
formation rate (solid line) or a sudden burst of star formation (short dashed
line). Also shown is the evolution assuming a metallicity $Z=0.05Z_\odot$ and
a constant star formation rate (long dashed line).
}
\label{fig:aeff}
\end{figure}

More critical is the value of $\alpha_S$, which sets the total number
of ionizing photons released by a source in terms of its specific
luminosity at the photoelectric threshold. The value for a QSO is
uncertain. Results based on {\it Hubble Space Telescope} ({\it HST})
data give $\alpha_S=1.76\pm0.12$ (Telfer \etal 2002). More recent
results based on {\it Far Ultraviolet Spectroscopic Explorer} ({\it
FUSE}) data give instead $\alpha_S=0.56^{+0.28}_{-0.38}$ (Scott \etal
2004). The origin of the discrepancy is unclear. The {\it FUSE} sample
is at lower redshift and the sources have lower luminosities, so the
difference may reflect a redshift and/or luminosity dependence. The
corrections due to intervening absorption by the IGM, however, are
significantly smaller for the {\it FUSE} sample than for the {\it HST}
sample, so that the {\it FUSE} result may be a more reliable estimate
of the spectral index of the mean QSO spectrum.

For a non-power-law source like a galaxy,
an effective spectral index $\alpha_S^{\rm eff}$ may be defined
according to
\begin{equation}
\alpha_S^{\rm eff}\equiv\frac{L_L}{h_P {\dot N_S}},
\label{eq:aeff}
\end{equation}
where $L_L$ is the specific luminosity of the source at the Lyman
edge. The effective spectral index is shown in Fig.~\ref{fig:aeff} for
starburst galaxies for both a fading sudden burst and for a constant
star formation rate, using {\sl Starburst99} (Leitherer \etal 1999),
updated with the stellar spectra of young stars from Smith, Norris \&
Crowther (2002). A Salpeter IMF is assumed and solar abundances.
While the spectral index of a fading burst quickly climbs to values
exceeding 10 (after $4\times10^7\,{\rm yrs}$), in the case of
continuous star formation, the effective spectral index instead
reaches an asymptotic value of $\alpha_S^{\rm eff}\approx2.3$. For
continuous star formation with Population II abundances
($Z=0.05Z_\odot$), the asymptotic value of the effective spectral
index is $\alpha_S^{\rm eff}\approx1.8$. This is comparable to the
effective spectral index for a blackbody spectrum with a temperature
of $T=5\times10^4\kel$, used by Stiavelli \etal to represent a galaxy
dominated by Population II stars. They represented Population III
stars (essentially no metals) by a blackbody with $T=10^5\kel$, for
which $\alpha_{\rm eff}\approx0.5$.

One implication of the rapidly climbing value of the effective index
in the case of a fading burst is that the ionizing efficiency of the
galaxy declines rapidly. If star formation in a bursting galaxy were
episodic, with a duty cycle of 50\% and a cycle time longer than
$2\times10^7\,{\rm yrs}$, the net contribution of the population to
the number of ionizing photons is reduced by 50\% from the case of
continuous star formation. This complication is not considered here.

Because of the uncertainty in the spectral index, three values are
considered here for the spectrum near the hydrogen Lyman edge:\ a hard
spectrum with $\alpha_S=0.5$ corresponding to a hard spectrum QSO or
Population III dominated galaxy, $\alpha_S=1.8$ corresponding to a
soft spectrum QSO or Population II dominated galaxy, and
$\alpha_S=2.3$ corresponding to a starburst with solar metallicity.

\begin{figure}
\begin{center}
\leavevmode 
\epsfxsize=3.3in
\epsfysize=3.3in
\epsfbox{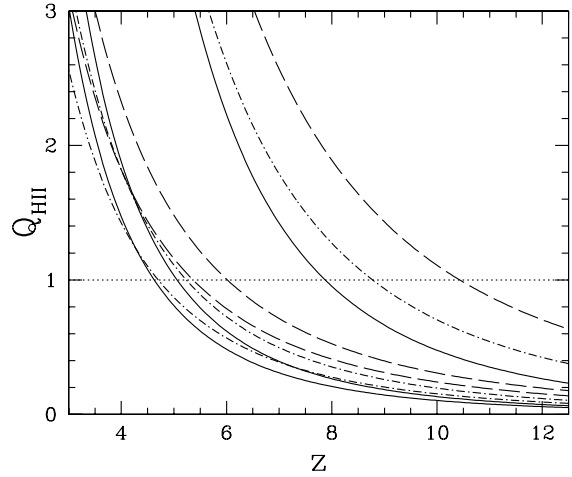}
\end{center}
\caption{The evolution of the ionization porosity parameter $Q_{\rm
HII}$ for hydrogen reionization based on the metagalactic ionizing
emissivity estimated from measurements of the mean transmitted \Lya
flux. The solid lines show the expected evolution assuming a source
spectral index of, from right to left, $\alpha_S=0.5$, 1.8 and
2.3. The dashed lines show the corresponding evolution for the
$2\sigma$ upper limits on the estimated emissivity. A clumping
factor of ${\cal C}=1$ is assumed for the above curves. Also shown are
the corresponding $2\sigma$ upper limits for ${\cal C}=3$ (dot-dashed
curves).
}
\label{fig:QHIIev}
\end{figure}

In Fig.~\ref{fig:QHIIev}, the evolution of the porosity parameter is
shown for hard and soft source spectra, and presuming the sources
turned on at very high redshifts ($z_i\rightarrow\infty$). Within the
observational uncertainties, the earliest epoch of reionization for
hard sources ($\alpha_S=0.5$), like QSOs or Pop~III stars, is $z_{\rm
ri}<11$ ($2\sigma$ upper limit). For soft sources ($\alpha=2.3$) like
starburst galaxies, $z_{\rm ri}<5.5\,(2\sigma)$.  Intermediate sources
with $\alpha_S=1.8$, like soft spectrum QSOs or Pop~II stars, may
ionize the IGM only at $z_{\rm ri}<6\,(2\sigma)$.  While hard spectrum
sources are well able to photoionize the IGM prior to $z=6$,
reionization by softer spectrum sources like soft spectrum QSOs or
starbursts (unless dominated by Pop~III stars), is more problematic.

The estimate provided neglects several factors which will further
delay the epoch of reionization. In particular, reionization will be
delayed if the galaxies turn on later (finite $z_i$) or if the
clumping factor is large. The source emissivity at the Lyman edge was
also equated to the total required metagalactic emissivity, neglecting
the additional ionizing photons arising from the IGM itself. Taking
this into account could reduce the estimated source emissivity by as
much as 50\% (see above), and so lower the overall predicted rate of
ionizing photons by this same amount, delaying reionization even
further.

The $2\sigma$ upper limits on $z_{\rm ri}$ for soft and intermediate
spectrum sources lie well below the {\it WMAP} $1\sigma$ and
$2\sigma$ lower limits of $z_{\rm ri}>14$ and 11, respectively (Kogut
\etal). The prediction for $\alpha_S=0.5$ is just barely compatible
with the {\it WMAP} $3\sigma$ lower limit of $z_{\rm ri}>8$. It should
be noted that Spergel \etal (2003) find the somewhat broader
reionization redshift constraint $z_{\rm ri}=17\pm4$ after combining
with other data sets. This still disagrees with the intermediate and
soft spectra predictions at the $3\sigma$ level, but agrees with the
prediction for $\alpha_S=0.5$ at the $\gsim2\sigma$ level. {\it The
results suggest that if the ionization of the IGM at $z<6$ is provided
by soft or intermediate spectrum sources, like starbursts dominated by
Pop~II stars, they are not members of the same population that
reionized the IGM, unless the population as a whole was a much more
prodigious source of ionizing radiation at earlier times ($z>6$).}
Sources with a hard spectrum ($\alpha_S=0.5$), like Pop~III stars or
hard spectrum QSOs, are preferred.

Gnedin (2004) similarly found that for his numerical simulations to
match the mean \Lya transmitted flux measurements at $z<6$, the epoch
of reionization (defined somewhat differently in terms of the
volume-weighted mean IGM ionization fraction), must have occurred at
$z_{\rm ri}\approx6$. The results here are much more generic and not
dependent on any particular prescription for cosmic star formation, as
in Gnedin's simulations. Gnedin finds that the predicted reionization
epoch is determined by a single parameter, the ultraviolet emissivity
parameter. Setting this parameter to match the mean \Lya optical depth
measurements at $z<6$ requires reionization to have occurred at
$z\approx6$. The one parameter nature of Gnedin's models appears to be
a consequence of the assumed nature of the ionizing sources formed as
part of the growth of cosmological structure. The analysis here shows
that reionization is determined by three parameters, expressed by
$A_S$, $\gamma$ and $\alpha_S$. The constraint on the reionization
epoch found here follows only from the assumption that the population
of sources which dominates the ionization background at $z<6$ is the
same population that reionized the IGM. The \Lya optical depth
measurements at $z<6$ are used to determine $A_S$ and $\gamma$,
leaving $\alpha_S$ free to set the reionization epoch.

In the next section, the QSO contributions to the metagalactic
ionizing background and to the total budget of ionizing photons are
estimated based on the QSO counts in recent surveys.

\section{The contribution of QSOs to the ionizing photon budget}
\label{sec:QSOs}
The QSO luminosity function for $z<2$ is well-established. The QSO
population detected by the 2dF (Croom \etal 2004) is well-fit at least
up to $z<2.1$ by a pure luminosity evolution (PLE) model with the
double power-law form (Boyle 1988):
\begin{equation}
\phi(M,z) = \frac{\phi^*}{10^{0.4(1-\beta_1)[M-M^*(z)]}+
10^{0.4(1-\beta_2)[M-M^*(z)]}},
\label{eq:phim}
\end{equation}
where $\phi^*$ is a fixed spatial normalization factor and $M^*(z)$ is
an evolving break magnitude.

For $z>3$, the model parameters are less clear since surveys have
generally only detected QSOs brighter than the break magnitude. The
most complete of these surveys is the QSO survey at $z>3.6$ carried
out as part of the SDSS (Fan \etal 2001, 2004). An exception is the
survey of Steidel \etal (2002), in which Active Galactic Nuclei (AGN)
dimmer than the break magnitude were discovered as part of a Lyman
break galaxy survey at $z\simeq3$.

In this section, the contribution of QSOs to the ionization rate of
the IGM over the redshift range $3<z<6$ is estimated based on the
findings of Fan \etal (2001, 2004), and as also constrained by the low
luminosity AGN data of Steidel \etal (2002). This is done for both
pure luminosity evolution and pure density evolution (PDE) models, for
which $M^*$ is fixed in Eq.~(\ref{eq:phim}) but $\phi^*$ is allowed to
evolve, as an attempt to bracket the uncertainty in the evolution.

The luminosity function parameters in Eq.~(\ref{eq:phim}) are
estimated using a maximum likelihood procedure. For any given set of
parameters, the predicted numbers of QSOs in a set of absolute
magnitude and redshift bins are given by $\mu_{ij}=\phi(M_i,z_j)V_{\rm
eff}^j(M_i) dM_i$ for magnitude bin $i$ and redshift bin $j$, where
$V_{\rm eff}^j(M_i)$ is the effective survey volume probed for QSOs
with absolute magnitudes within the bin centred on $M_i$ and redshifts
in the bin centred on $z_j$. All absolute magnitudes are taken at the
published values at 1450A, and a flat cosmological model with
$\Omega_{\rm M} =0.35$ and $h=0.65$ is used for the fits for
consistency with Fan \etal (2001, 2004). (The magnitudes and effective
volumes from Steidel \etal (2002) and Hunt \etal are transformed to
this cosmology.) The likelihood of the model compared with the actual
numbers $N_{ij}$ of QSOs detected is then given by
\begin{equation}
{\cal L} = \prod_{ij}\frac{e^{-\mu_{ij}}\mu_{ij}^{N_{ij}}}{N_{ij}!},
\label{eq:likelihood}
\end{equation}
where Poisson probabilities are required because of the low number of objects
per bin.

The likelihoods are computed on a multi-dimensional (linear) grid in
$\beta_1$, $\beta_2$, $\phi^*$ and $M^*$. (This corresponds to
assuming a uniform prior for each parameter.) Here $\beta_1$ is
assumed to correspond to the low luminosity end of the QSO luminosity
function and $\beta_2$ to the high end. For the PLE models, the same
$\phi^*$ is used for all redshift bins, but $M^*$ is stepped over
separately for each redshift bin. Conversely, for the PDE models the
same $M^*$ is adopted for all redshifts, but $\phi^*$ is allowed to
vary for each redshift bin. Over $10^9$ models are constructed for the
PLE case and $10^8$ for the PDE case.

\begin{figure}
\begin{center}
\leavevmode 
\epsfxsize=3.3in
\epsfysize=3.3in
\epsfbox{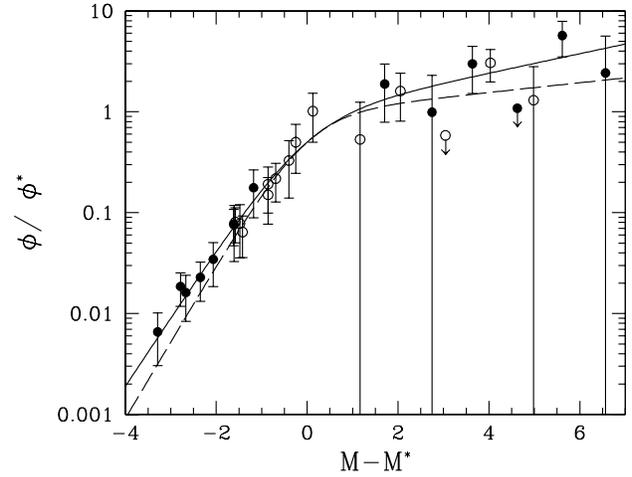}
\end{center}
\caption{The maximum likelihood fits to the QSO luminosity function
for the PLE (solid line) and PDE (dashed line) models. The measured
luminosity function values have been renormalized by $\phi^*$ and
$M^*$ for the PLE model (solid points) and the PDE model (open
points). The error bars are Poisson, based on the maximum likelihood
fits.}
\label{fig:phifits}
\end{figure}

The parameter values for the maximum likelihood fit of the PLE model
are $\beta_1=1.24$, $\beta_2=2.70$ and
$\phi^*=3.25\times10^{-7}\,{\rm Mpc^{-3}\,mag^{-1}}$ (comoving). The
values for $M^*(z)$ are given in Table~\ref{tab:phifits}. The
parameters for the PDE fit are $\beta_1=1.12$, $\beta_2=2.90$ and
$M^*=-25.60$. The values for $\phi^*(z)$ are given in
Table~\ref{tab:phifits}.

The goodness-of-fit of the maximum likelihood models is estimated
using the $\chi^2$-test for the predicted vs detected number of QSO
sources for all 15 available bins in redshift and magnitude (one of
which is a cumulative magnitude bin), and allowing for 7 degrees of
freedom (after accounting for the 8 model parameters). For the PLE
model, $\chi^2=6.8$ and $p(\chi^2>6.8)=0.45$. For the PDE model,
$\chi^2=8.5$ and $p(\chi^2>8.5)=0.29$. Both fits are acceptable.  The
maximum likelihood fits are shown in Fig.~\ref{fig:phifits}, along
with the (renormalized) measured values of the luminosity function and
their Poisson errors, which are based on the fit models.

The maximum likelihood parameter values, each marginalised over all
other model parameters, and with their equivalent $1\sigma$ error
ranges, are, for the PLE model, $\beta_1=1.24^{+0.12}_{-0.12}$,
$\beta_2=2.2^{+0.3}_{-0.2}$ and $\phi^*=2.8^{+3.0}_{-1.8}
\times10^{-7}\,{\rm Mpc^{-3}\,mag^{-1}}$ (comoving). The marginalised
values for $M^*(z)$ are given in Table~\ref{tab:phifits}. For the PDE
model, the values are $\beta_1<1.1$, $\beta_2=2.8^{+0.4}_{-0.5}$ and
$M^*=-25.4^{+0.8}_{-0.6}$. The marginalised values for $\phi^*(z)$ are
given in Table~\ref{tab:phifits}. The upper limit on $\beta_1$ results
since smaller values were not considered.

The results on $\beta_1$ and $\beta_2$ agree well with the findings of
Hunt \etal and Fan \etal (2001, 2004). Hunt \etal obtain
$\beta_1=1.24\pm0.07$ at $z\approx3$. Fan \etal (2001) obtain, for
$3.6<z<5.0$, $\beta_2=2.58\pm0.23$, while at $z>5.7$, Fan \etal (2004)
obtain $\beta_2=3.2^{+0.8}_{-0.7}$. (All errors are $1\sigma$.) The
PLE model estimate of $\beta_2$ found here disagrees with the $z>5.7$
value at the 98\% confidence level, suggesting some departure from
PLE, but the evidence is not conclusive.

For $0.4<z<2.1$, Croom \etal obtain $\beta_1=1.0\pm0.1$ and
$\beta_2=3.25\pm0.05$ for a PLE model with a break magnitude varying
as an exponential of lookback time. (The results for a quadratic
redshift dependence are similar.) There is again only marginal
agreement with $\beta_2$. There is greater disagreement over the
normalizations. Croom \etal report $\phi^*_B=1.8\times10^{-6}\,{\rm
Mpc^{-3}\,mag^{-1}}$ for the $B-$band luminosity function. This value
(including an adjustment for the small differences in assumed
cosmologies), exceeds the values for $\phi_*$ found for the PLE model
here by a factor of $6^{+10}_{-3}$, significant at the $5\sigma$
level. It should, however, be noted that the normalizations may only
be fairly compared provided the ratio of $B-$band to 1450A
luminosities was itself luminosity-independent. A steep correlation
between spectral slope and $B-$band magnitude for QSOs could in
principle account for the difference in normalizations. Currently
there is no evidence for a correlation sufficiently steep. It thus
appears there has been substantial density evolution from $z\approx3$
to $z\lsim2$.

\begin{figure}
\begin{center}
\leavevmode 
\epsfxsize=3.3in
\epsfysize=3.3in
\epsfbox{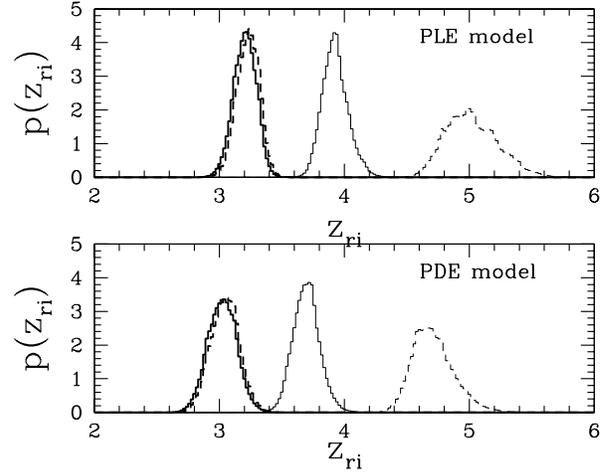}
\end{center}
\caption{The probability distributions for the epoch of reionization
for hydrogen (solid lines) and helium (dashed lines) assuming a QSO
spectral index of $\alpha_S=1.8$ (heavy lines) or 0.5 (light lines).}
\label{fig:pzri}
\end{figure}

For each set of luminosity function parameters, the integrated
emissivity values at the Lyman edge for the PLE and PDE models are
computed from $\epsilon_L=\int \phi(M) L_L(M) dM$, where $L_L(M)$ is
the frequency specific QSO luminosity at the Lyman edge corresponding
to a given absolute magnitude $M$ at 1450A, and assuming
$L_\nu\propto\nu^{-0.5}$ between 1450A and the Lyman break. The
resulting marginalised maximum likelihood emissivity values
(transformed to a flat cosmology with $\Omega_{\rm M}=0.3$ and $h=0.7$), are
shown in Fig.~\ref{fig:emiss} for the PLE model (filled squares) and
the PDE model (open squares), along with their $1\sigma$ errors. The
corresponding marginalised estimates for the ionizing photon
production rates ($\alpha_S {\dot N}_S=\epsilon_L/h_P$) are shown in
Fig.~\ref{fig:Nphot}.

As shown in Fig.~\ref{fig:emiss}, for $z\lsim3$, all of the required
emissivity is accounted for by QSOs. At $z>4.5$, however, it appears
QSOs are inadequate sources of ionizing photons, falling short by
factors of at least $2-3$, depending on the amount of diffuse
radiation. Additional sources are required to make up the
difference. Within the uncertainties, the QSO deficit could be as high
as a factor of 30 by $z\approx5.5$.

The deficit of ionizing photons from QSOs is apparent in
Fig.~\ref{fig:Nphot} as well. For $z>5.5$, QSOs fall short of the
ionizing photon production rate required for consistency with the
measured mean transmitted \Lya flux values (solid and long-dashed
lines) by at least a factor of 5. The $1\sigma$ upper limit on the QSO
rate for the PLE case at $z=5.7$, however, comes close to the bare
minimum rate required to ionize all the baryons over a Hubble time,
assuming a hard ($\alpha_S=0.5$) source spectrum. While QSOs could
not have ionized the IGM acting alone, they still may have provided a
non-negligible contribution of ionizing photons during the
reionization epoch.

\section{Discussion}
\label{sec:discussion}
Any assessment of the contribution of a population of sources to the
ionizing photon budget of the IGM depends on several unknown
factors. Principal among these are the spectral index $\alpha_{\rm
MG}$ of the UV background, the contribution of Lyman Limit Systems to
the attenuation length at the Lyman edge, the amount of diffuse
emission arising from the IGM, and the clumping factor of the IGM. In
this section, the role each of these factors may play is discussed.

\subsection{The sources of ionization of the IGM}
\label{subsec:ionization}
Comparison of the results of numerical simulations of the IGM with
measurements of the mean transmitted \Lya flux through the IGM
constrains the ionization rate per hydrogen atom (or per \HeII atom
for \HeII \Lya). This is related to the emissivity by a factor of
$3+\alpha_{\rm MG}$ (Eq.~\ref{eq:Geps}). The results of Haardt \&
Madau suggest $\alpha_{\rm MG} \approx1$ shortward of the hydrogen
Lyman edge, but this is due in large part to the contribution of
redshifted \HeII recombinant \Lya and two-photon emission. If most of
the helium is still in the form of \HeII, the spectrum may be harder,
possibly even with $\alpha_{\rm MG}<0$. In this case, the required
emissivity could be reduced by as much as a factor of two, bringing it
much more in line with the expected QSO emissivity.

Meiksin \& White (2004) appealed to the uncertainty in the values of
$\alpha_{\rm MG}$ and the QSO break magnitude ($M^*$), adjusting these
to allow QSOs to dominate the UV ionizing background at high
redshifts. The assessment here does not rule out this
possibility. Since QSOs acting alone would not be able to photoionize
the IGM by $z=6$, however, it seems likely the UV background will have
a large contribution from other sources, such as galaxies. (Although
this need not be the case if reionization occurred at $z>>6$.)  While
this would suppress the correlations in the UV background predicted by
Meiksin \& White (2004), and the predicted enhancement of the \Lya
flux correlations in the IGM, the suppression may be by no more than a
factor of a few. In this case, the contribution of QSOs to the UV
background may still be detectable through the fluctuations they
induce.

The ionization rate predicted for a given emissivity also depends on
the attenuation length $r_0$. The values used here were computed
self-consistently from numerical simulations, but these simulations do
not include radiative transfer and therefore are not able to
accurately predict the contribution of Lyman Limit Systems. These are
accounted for by direct observations for $z\lsim4$, but their
abundances are not known at higher redshifts. The attenuation length
may be $\sim30$\% smaller than the values adopted here, resulting in
an increase in the required emissivity by the same amount. This would
result in a further reduction in the fraction of the ionizing
background provided by QSOs. A similar effect may arise if the
emissivity of the sources rapidly declines with redshift, in which
case the effective path length over which the emissivity is integrated
to obtain the ionization rate may be comparable or even shorter than
the attenuation length, which also would permit a higher emissivity at
a given redshift (Meiksin \& White 2003).

If a substantial fraction of the required emissivity ($\sim~50$\%)
arises from diffuse recombination radiation within the IGM, then the
requirements for the {\it source} emissivity are reduced by the same
amount. The fractional contribution of QSOs to the required source
emissivity would then increase. The difference between the required
source emissivity and the predicted emissivity from QSOs for the PLE
and PDE cases is shown in Fig.~\ref{fig:emiss}, assuming the diffuse
emission from the IGM contributes at 50\% the source level. QSOs
provide more than half of the required source emissivity up to
$z\lsim4.5$, and less beyond. The curves showing the difference
describe the emissivity contributed by other sources, presumably
galaxies.

\subsection{The sources of reionization of the IGM}
\label{subsec:reionization}
The reionization redshifts estimated here rest on the assumption that
the population of sources which dominates the UV ionizing background
at $z<6$ is the same population responsible for reionization. This is
to be expected if reionization occurred near $z\approx6$, as argued by
Becker \etal, but there is nothing to preclude a much earlier
reionization epoch ($z>>6$), or even a series of several reionization
epochs followed by recombination. On the other hand, the mild
evolution in the emissivity required to match the measured \Lya
optical depths at $z<6$ suggests there was no abrupt change in the
nature of the sources dominating the UV background at $z\approx6$, so
that the emissivity may be reasonably extrapolated to higher
redshifts.

The estimated reionization redshifts were all upper limits, neglecting
any excess clumping (${\cal C}>1$) in the reionized gas. If the
clumping factor is moderate (${\cal C}\approx2-4$), as suggested by
recent numerical computations, then recombinations may not delay the
reionization epoch very significantly. The evolution of the porosity
for ${\cal C}=3$, assuming the $2\sigma$ upper limit on the
emissivity, is shown in Fig.~\ref{fig:QHIIev}. For $\alpha_S=2.3$ and
1.8, reionization now occurs well below $z=6$. For $\alpha_S=0.5$, the
$2\sigma$ upper limit of $z_{\rm ri}<9$ is still compatible with the
SDSS data (Becker \etal), but barely with the {\it WMAP} $2\sigma$
lower limit of Spergel \etal The expected value of the reionization
redshift for $\alpha_S=0.5$ and ${\cal C}=3$ is $z_{\rm ri}=7.1$,
which is compatible with the {\it WMAP} $2.5\sigma$ lower limit of
Spergel \etal For higher values of the clumping factor (${\cal
C}\approx5-10$), reionization for the various spectral index cases
considered here would occur substantially later.

If the reionization of minihalos doubles the reionization requirements
(Shapiro \etal 2004, Iliev \etal), then reionization for
$\alpha_S=0.5$ is not expected to occur until $z_{\rm
ri}\lsim6.2$. While this is consistent with the data of Becker \etal,
it is consistent with the {\it WMAP} lower limit on $z_{\rm ri}$ of
Spergel \etal at only the $2.7\sigma$ level.

The estimates here assumed the sources turned on at infinitely high
redshifts ($z_i>>z_{\rm ri}$). Reionization would occur somewhat later
for finite values of $z_i$.

The estimates for the reionization epoch were also based on the
assumption that all the required emissivity was provided by the
sources, while in fact diffuse emission from the IGM could contribute
an additional $\sim50$\%. In this case, the expected source
contribution would be reduced by 50\%, and $(1+z_{\rm ri})$ reduced by
about 15\%. For $\alpha_S=0.5$, this would push the $2\sigma$ upper
limit on $z_{\rm ri}$ down from 11 to 9.

Reionization may have occurred earlier if the emissivity values used
here are underestimated, as would be the case if the attenuation
length of Lyman photons were shorter than the values used here (see
\S~\ref{subsec:ionization} above). Even a doubling of the
emissivities, however, would increase the expected values of
$(1+z_{\rm ri})$ by only about 20\%, or $z_{\rm ri}\approx10$ for
$\alpha_S=0.5$. For $\alpha_S=1.8$ and 2.3, the expected redshifts for
the reionization epoch may be increased to $z_{\rm ri}\approx6$, still
too low to match the {\it WMAP} constraints, but consistent with the
SDSS \Lya optical depth measurements. At much higher redshifts, the
reionization epoch will be limited by the effect of recombinations
even for small clumping factors.

In the scenarios above, the expected redshift of \HeII reionization is
always above $z_{\rm ri}>11$ for $\alpha_S=0.5$ (even for ${\cal
C}=3$), so that helium may have been ionized very early if the source
spectra continue as hard power laws shortward of the \HeII Lyman
edge. If the spectrum softens to $\alpha_S>1.8$ shortward of the \HeII
Lyman edge, however, \HeII reionization would not have occurred prior
to $z_{\rm ri}\lsim6$ (except if the total emissivity is doubled), but
generally prior to $z_{\rm ri}\gsim3$, unless \HeII ionizing photons
are produced at a negligible rate, as is expected if the ionizing
sources are star forming galaxies (Smith \etal).

QSOs are found to provide too few ionizing photons at $z>5$ to
reionize the hydrogen in the IGM. This may be quantified in terms of
the ionization porosity. Because of the rapid decline in the photon
production rate, Eq.~(\ref{eq:QHIIevrec}) no longer applies. Instead,
the decline in Fig.~\ref{fig:Nphot} is better modelled as an
exponential in redshift:\ $\alpha_S{\dot N}_S=A_Se^{-\gamma z}$. The
solution of Eq.~(\ref{eq:QHII}) (with recombinations neglected) is
\begin{equation}
Q_{\rm HII}=\frac{{\dot N}_S(z)t_{\rm H}(z)}{n_{\rm H}(0)}
\left[1-2y+2 e^y y^{3/2}\Gamma\left(\frac{1}{2},y\right)\right],
\label{eq:QHIIeev}
\end{equation}
where $y=\gamma(1+z)$, and $z_i\rightarrow\infty$ is assumed. A similar
expression follows for $Q_{\rm HeIII}$ on replacing $n_{\rm H}(0)$ by
$n_{\rm He}(0)$ and ${\dot N}_S$ by ${\dot N}_S/4^{\alpha_S}$.

Because of the broad, non-gaussian probability distribution for ${\dot
N}_S$ from QSOs, the probability distribution for the reionization
redshift is determined from a maximum likelihood estimate using the
probability distributions found in \S~\ref{sec:QSOs} for the QSO
emissivity.  For the PLE case, the parameters of the maximum
likelihood model are $A_S=5.7\times10^{52}\,{\rm ph\,s^{-1}}$ and
$\gamma=1.26$. For the PDE case, $A_S=6.1\times10^{52}\,{\rm
ph\,s^{-1}}$ and $\gamma=1.37$. The marginalised maximum likelihood
values for the parameters are, for the PLE case,
$A_S=5.2^{+6.2}_{-3.1}\times10^{52}\,{\rm ph\,s^{-1}}$ and
$\gamma=1.65+0.10\log(A_S/5.2\times10^{52})$. For the PDE case,
$A_S=5.7^{+7.4}_{-3.3}\times10^{52}\,{\rm ph\,s^{-1}}$ and
$\gamma=1.74+0.12\log(A_S/5.7\times10^{52})$.

The resulting probability distributions for the reionization
redshifts, both for \HII and \HeIII, are shown for the PLE and PDE
cases in Fig.~\ref{fig:pzri}. Results for both $\alpha_S=1.8$ and 0.5
are provided. For $\alpha_S=1.8$, the hydrogen and helium reionization
redshifts nearly coincide (since the coefficients for $Q_{\rm HII}$
and $Q_{\rm HeIII}$ are nearly identical for this value of the
spectral index), and correspond to reionization redshifts $z_{\rm
ri}\approx3$. While inconsistent with the \HI data, measurements of
the mean transmitted \HeII \Lya flux reveal the absence of a
measurable flux by $z\approx3.5$ (Zheng \etal 2003). Any future
detections of a finite \HeII \Lya optical depth at $z>3.5$ would
suggest the QSO spectra are harder than $\alpha_S=1.8$ shortward of
the \HeII Lyman edge. For $\alpha_S=0.5$, helium reionization may
occur as early as $z\approx5$ by QSOs alone. By contrast, even for
this hard a spectrum, QSOs alone are unable to reionize the hydrogen
in the IGM any earlier than $z\approx4$.

If helium was not fully ionized until $z\approx3$, then the UV
background may have undergone a second change in its spectral shape
shortward of the hydrogen Lyman edge due to the release of
(redshifted) recombinant \HeII photons and two-photon emission (Haardt
\& Madau). (The first change may have been at $z<4.5$ when QSOs began
to dominate the source emissivity.) There would of course also be a
marked change shortward of the \HeII Lyman edge. In addition to the
indirect effect on the temperature of the IGM, these changes may be
detectable through their effects on the relative abundances of
ionization states of metal ions (eg, \SiIV/\CIV), for which there is
some evidence at $z\approx 3.1$ (Songaila \& Cowie 1996).

\subsection{Did low luminosity AGN reionize the IGM?}
\label{subsec:reionAGN}

The implication that the reionization sources had hard spectra
($\alpha_S=0.5$) would suggest that star forming galaxies (or star
clusters) dominated by Pop~III stars are the only viable candidates
for reionizing the IGM. QSOs, the only other hard spectra sources
known, decline too rapidly in their number at high
redshifts. Reionization soley by Pop~III stars, however, has its own
difficulties, as the stars would rapidly pollute their environment
with metals, so that reionization would be accomplished predominantly
by Pop~II stars (Schneider \etal 2002; Ricotti \& Ostriker 2004). As
shown above, Pop~II stars are just barely able to reionize the IGM by
$z=6$ without violating the emissivity requirements needed to match
the measurements of the mean transmitted \Lya flux at $z<6$. They fall
well short of the {\it WMAP} reionization requirement. Pop~III stars
(or Pop~II stars) could only have reionized the IGM if their
emissivity rapidly declined from the time of reionization to $z<6$.

Madau \etal (2004) discuss miniquasars as an alternative reionization
source, arguing that they may arise naturally in the same systems
forming Pop~III stars through accretion onto black holes formed from
collapsed massive stars. Dijkstra, Haiman \& Loeb (2004), however,
claim that this and any scenario in which reionization was performed
by quasar-like objects are excluded by constraints from the soft X-ray
background. In this section, the possibility that low luminosity AGN
reionized the IGM is re-explored.

From an energetic perspective, the requirements that galaxies
harbouring active nuclei at high redshifts must satisfy for
reionization are readily met. A black hole of mass $10^{6.5}\,M_\odot$
would shine with an Eddington luminosity of $L_{\rm
Edd}\simeq5\times10^{44}\,{\rm ergs\, s^{-1}}$, producing a specific
luminosity at the Lyman edge of $L_L\sim 0.1L_{\rm Edd}/\nu_L
\approx10^{28}\,{\rm ergs\,s^{-1}\,Hz^{-1}}$ (allowing for a
contribution near the Lyman edge to the bolometric luminosity of
10\%). At $z=6$, Eq.~\ref{eq:emiss} gives a total (comoving)
emissivity of $\epsilon_L\approx3\times10^{24}\,{\rm
ergs\,s^{-}\,Hz^{-1}\, Mpc^{-3}}$. The corresponding comoving space
density of galaxies harbouring such black holes is then
$n_g\sim2\times10^{-4}\,{\rm Mpc^{-3}}$, only a small percentage of
galaxies today. The magnitude at 1450A would then be
$M_{1450}\approx-19$ (assuming $f_\nu\sim\nu^{-0.5}$). This
corresponds to a $z'$ magnitude of $\gsim27$ at $z\sim6$ and $\gsim28$
at $z\sim8$. The number of sources per unit redshift per square
arcminute at $6<z<8$ would then be $dN/dz\approx1/2$. It is possible
that as many as a dozen of such objects appear as $i'$-dropouts in the
UDF (see, eg, Bunker \etal 2004). At such low luminosities, however,
they would have escaped detection in the GOODS fields and in the
$z\approx3$ survey of Steidel \etal (2002). It does imply a sharp turn
up at faint magnitudes in the number of sources compared with the Hunt
\etal luminosity function at $M_{1450}=-20$. The comoving density of
the objects, however, may have declined substantially by $z\approx3$,
as they are no longer required to maintain the ionization of the IGM,
since the known QSOs are adequate (Fig.~\ref{fig:emiss}). The excess
towards fainter magnitudes may then be only mild.

The low luminosities and abundance of the objects also place a minimal
demand on the number of ionizing photons required per hydrogen atom to
reionize the IGM. The typical (proper) distance between the sources at
$z=11$ is $\sim1$~Mpc, corresponding to a flux of ionizing photons of
$F\sim10^{53}\,{\rm ph\,s^{-1}\,Mpc^{-2}}$, or $F_0=0.01-0.1$ in the
units of Iliev \etal Such low fluxes require only 1--2 ionizing
photons per hydrogen atom to photoionize a minihalo (Iliev \etal) as a
consequence of photoevaporation.

The objects would also be consistent with constraints imposed by the
soft X-ray background. The estimate of Dijkstra \etal was based on
assuming a mean requirement of 10 ionizing photons per hydrogen atom
to reionize the IGM. They then argued that any quasar-like sources
which produced sufficient photons to reionize the IGM would exceed by
a factor of about 4--6 the component of the residual soft X-ray
background unaccounted for by discrete sources. As this factor scales
linearly with the number of ionizing photons required per atom, a
requirement of only 2 photons per atom would bring the level into
agreement with the residual background. (This argument may apply to
miniquasars as well.)

Even without appealing to the decrease in the number of required
ionizing photons per atom, however, the soft X-ray background
constraint may be avoided because of the presumed hard spectrum
($\alpha_S=0.5$) of the sources. Dijkstra \etal based their estimates
for QSOs on the spectrum of Sazonov, Ostriker \& Sunyaev (2004), who
assumed a mean QSO energy spectrum varying as (their eq.~[14])
$\langle F(E)\rangle\sim E^{-0.6}$ for $1\,{\rm eV}\leq E<10\,{\rm
eV}$, and $\langle F(E)\rangle \sim E^{-1.7}e^{E/2\,{\rm keV}}$ for
$10\,{\rm eV}\leq E<2\,{\rm keV}$, based on the {\it HST} data of
Telfer \etal Adopting $\alpha_S=0.5$ at $E>10\,{\rm eV}$, however, can
increase the number of ionizing photons while leaving the X-ray
constraints unchanged. For instance, adopting the form (with $E$ in
units of keV)
\begin{equation}
\langle F_E\rangle=210\cases{0.63E^{-0.6} & $0.001\leq E<0.01$,\cr
E^{-0.5}e^{-E/0.4} & $0.01\leq E<2$,\cr}
\label{eq:FE}
\end{equation}
preserves the optical to X-ray spectral index $\alpha_{ox}$ between
2500A and 2~keV (as well as the normalization at 2~keV), but produces
3 times as many hydrogen ionizing photons as the corresponding model
of Sazonov \etal (It also predicts $L_L\simeq0.1L_{\rm bol}/ \nu_L$,
used above to relate $L_L$ to $L_{\rm Edd}$.)  This again would bring
the predicted contribution of the sources to the soft X-ray background
into agreement with the residual component. It thus appears that hard
spectrum quasar-like objects may reionize the IGM without exceeding
the limits of the soft X-ray background limits.

\section{Conclusions}

A relation between the mean transmitted \Lya flux and ionization
porosity of the IGM is exploited to obtain generic limits on the epoch
of reionization. Under the assumption that the sources which dominate
the UV ionization background at $z<6$ are members of the same
population of sources that reionized the IGM, it is shown that the
epoch of reionization must have occurred at $z_{\rm ri}<11$ ($2\sigma$
upper limit) for hard spectrum ($\alpha_S=0.5$) sources like QSOs, AGN
and Pop~III dominated star forming galaxies, $z_{\rm ri}<6$
($2\sigma$) for moderate spectrum ($\alpha_S=1.8$) sources like soft
spectrum QSOs, AGN and Pop~II dominated star forming galaxies, and
$z_{\rm ri}<5.5$ ($2\sigma$) for soft spectrum ($\alpha_S=2.3$)
sources like star forming galaxies with solar abundances.

After reionization, it is possible that no single population of
sources dominated the UV ionizing background down to
$z\approx2$. Instead the UV ionizing background may have undergone
multiple qualitative changes in its spectral shape as the nature of
the dominant ionizing sources changed. An assessment of the
contribution of QSOs to the budget of ionizing photons based on recent
high redshift QSO surveys suggest that QSOs alone are adequate for
providing the ionization rate required to reproduce the mean \Lya
optical depth measurements at $z\lsim3$. QSOs produce at least half of
the required emissivity at $z\approx3.5$, but fall short by a factor
of at least $2-3$ for $4.5<z<6$. As a consequence, the UV metagalactic
background shortward of the Lyman edge may have undergone a
qualitative change in its spectral shape in the interval
$3.5<z<4.5$. QSOs produce too few hydrogen ionizing photons to
reionize the IGM any earlier than $z_{\rm ri}\approx4$, even assuming
a hard spectral index ($\alpha_S=0.5$). For a moderate spectral index
($\alpha_S=1.8$), QSOs alone are unable to ionize the helium in the
IGM any earlier than $z\approx3$, but helium reionization may occur as
early as $z\approx5$ for QSOs with a hard spectral index
($\alpha_S=0.5$) (unless the clumping factor is large). If QSOs do not
ionize \HeII until $z\approx3$, then a second qualitative change will
result in the spectral shape of the UV background.

Both PLE and PDE models are found to provide acceptable fits to the
QSO counts over the redshift range $3<z<6$. Comparison with PLE models
for $z<2$, however, suggests substantial density evolution in the QSO
luminosity function between $z>3$ and $z\lsim2$. Predicted break
magnitudes for the models at $z>3$ suggest that the breaks would be
clearly detectable in the PDE case for a survey one magnitude deeper
than the SDSS QSO survey and three magnitudes deeper for the PLE case.

Consistency with the {\it WMAP} $2\sigma$ lower limit of $z_{\rm
ri}>11$ (Kogut \etal) suggests that the sources which reionized the
IGM must have had hard spectra ($\alpha_S\approx0.5$), like hard
spectra QSOs or AGN, or Pop~III stars, unless reionization occurred
very early ($z>>6$), in which case the ionization state of the IGM at
$z<6$ would not be expected to yield any constraints on the nature of
the reionization sources. An assessment of the contribution of
ionizing photons from QSOs based on recent surveys shows that they
fall short of the minimal requirement for reionizing the IGM by a
factor of several. This suggests that either:\ 1.\ galaxies were much
more prodigious sources of ionizing photons at $z>6$ than at $z<6$,
either because of a higher star formation rate or a higher escape
fraction of ionizing photons, 2.\ star formation in galaxies at $z>6$
was dominated by the production of Pop~III stars, producing a hard
spectrum, 3.\ at least a small fraction of galaxies harboured a low
luminosity AGN at $z>6$, so that they were hard spectrum sources at
these epochs, or 4.\ reionization was produced by an as yet
unidentified population of sources (such as miniquasars).

Another possibility is that numerical simulations of the structure of
the IGM at high redshifts are seriously in error. The high precision
agreement with measurements of the \Lya forest in spectra at moderate
redshifts ($z<4$) (Meiksin, Bryan \& Machacek 2001) does not preclude
the possibility that energetic events such as galactic winds or QSOs
at higher redshifts substantially disturbed the gas at earlier times,
so that current simulations do not provide valid constraints on the
source emissivity required to match the mean \Lya transmitted flux
measurements at $z>5$. Currently there is no direct evidence that this
may be the case.

\bigskip

\begin{table}
\begin{center}
\begin{tabular}{|c|c|c|c|c|} \hline
Redshift & PLE:\ $M^*$ & PDE:\ $\phi^*$& PLE:\ $M^*$ & PDE:\ $\phi^*$\\
&(ML)&(ML)&(margin.)&(margin.)\\
\hline 
\hline
$z=3$ &$-27.2$ & $6.0$ & $-26.2^{+1.6}_{-1.8}$& $6.1^{+3.2}_{-2.7}$ \\
\hline
$3.6<z<3.9$ & $-24.7$ & $1.2$ & $-24.0^{+0.7}_{-0.8}$& $1.1^{+5.0}_{-0.3}$ \\
\hline
$3.9<z<4.4$ & $-24.4$ & $0.74$ & $-23.6^{+0.7}_{-0.9}$& $1.3^{+2.4}_{-0.8}$ \\
\hline
$4.4<z<5.0$ & $-23.8$ & $0.27$ & $-22.9^{+0.9}_{-1.0}$& $0.43^{+1.23}_{-0.16}$ \\
\hline
$z>5.7$ & $-22.9$ & $0.08$ & $-21.4^{+0.9}_{-1.4}$& $0.09^{+0.34}_{-0.03}$ \\
\hline

\end{tabular}
\end{center}
\caption{Redshift dependent parameters for the QSO luminosity function
maximum likelihood estimates. The two leftmost columns of values provide
the parameters
for the maximum likelihood model for the PLE and PDE cases, as indicated.
The two rightmost columns provide the marginalised
parameter values, and their $1\sigma$ errors, for the PLE and PDE cases.
All magnitudes are at 1450A. The normalization $\phi^*(z)$ is in units of
$10^{-7}\,{\rm Mpc^{-3}\, mag^{-1}}$ (comoving). The fits are done for a flat
cosmology with $\Omega_{\rm M}=0.35$ and $h=0.65$.
}
\label{tab:phifits}
\end{table}

%%%%%%%%%%%%%%%%%%%%%%%%%%%%%%%%%

\end{document}